# Transcriptional Regulation by the Numbers 2: Applications

Lacramioara Bintu[1], Nicolas E. Buchler[2], Hernan G. Garcia[3], Ulrich Gerland[4], Terence Hwa[5], Jané Kondev[1], Thomas Kuhlman[5] and Rob Phillips[6]

[1]*Physics Department, Brandeis University, Waltham, MA 02454*

[2]*Center for Studies in Physics and Biology, The Rockefeller University, New York, NY 10021*

[3]*Department of Physics, California Institute of Technology, Pasadena, CA 91125*

[4]*Physics Department and CENS, Ludwig-Maximilians University, Munich, Germany*

[5]*Physics Department and Center for Theoretical Biological Physics, University of California at San Diego, La Jolla, CA 92093-0374*

[6]*Division of Engineering and Applied Science and Kavli Nanoscience Institute, California Institute of Technology, Pasadena, CA 91125*

**ABSTRACT**

With the increasing amount of experimental data on gene expression and regulation, there is a growing need for quantitative models to describe the data and relate them to the different contexts. The thermodynamic models reviewed in the preceding paper provide a useful framework for the quantitative analysis of bacterial transcription regulation. We review a number of well-characterized bacterial promoters that are regulated by one or two species of transcription factors, and apply the thermodynamic framework to these promoters. We show that the framework allows one to quantify vastly different forms of gene expression using a few parameters. As such, it provides a compact description useful for higher-level studies, e.g., of genetic networks, without the need to invoke the biochemical details of every component. Moreover, it can be used to generate hypotheses on the likely mechanisms of transcriptional control.

**Introduction**

Biology is undergoing a transformation from a component-centric focus on the parts towards a *system-level* focus on how a limited number of parts work together to perform complex functions. For gene regulation, this theme has been discussed extensively in the context of simple genetic circuits [1,2,3,4] as well as complex, developmental networks [5]. The functional properties of a genetic circuit often depend critically on the degree of cooperativity in the interactions between the molecular components [6]. For gene regulation, this cooperativity is dictated to a large extent by the architecture of the cis-regulatory region [7] and the specific mechanism of transcription activation or repression [8] mediated through interactions among various transcription factors (TF) and the RNA polymerase complex (RNAP). Often, even qualitative features of a gene circuit (e.g., whether a circuit can be bistable or can spontaneously oscillate) cannot be determined without *quantitative* knowledge of the transcription regulation of key genes in the circuit [3].

Quantifying the level of gene expression from a promoter starting from the underlying biochemistry/biophysics is a difficult task, due most notably to ignorance of many biochemical parameters, especially their relevant *in vivo* values. On the other hand, the thermodynamic model reviewed in the preceding article [9] yields several general mathematical forms for the dependence of the fold-change in gene expression on the concentration(s) of the TF(s) regulating transcription. These general forms contain only a few parameters characterizing the *effective* interactions between the molecular players. Thus, from a practical standpoint, it seems expedient to quantify the transcription regulation of a gene by fitting expression data to the appropriate model function to obtain effective parameters that best describe the promoter [10,11]. This procedure may be useful even when the simplifying assumptions made by the thermodynamic models are not satisfied [9]. By analyzing gene expression data within the thermodynamic framework, one can elucidate whether an assumed set of interactions between TFs and RNAP can consistently explain the data. Failure of the analysis can suggest important missing ingredients, such as unknown mechanisms of cooperativity, while success can lead to predictions for new experiments, e.g., how operator deletion would affect gene expression.

There has been much recent progress in understanding the mechanistic aspect of bacterial gene regulation [8]. However, the systematic *quantification* of gene expression is still in its infancy. In this paper, we review a number of experimentally characterized cis-regulatory systems in bacteria, and provide for each case what we believe to be the most appropriate form for the dependence of the fold-change in promoter activity on the TF concentration(s). For each system, we show graphically how the expected form depends on the effective parameters. We hope to demonstrate how the thermodynamic models can provide a direct link between the arrangements of interactions in a promoter region and the quantitative characteristics of gene expression.

**Quantitative Characteristics of Activation and Repression**

Our quantitative discussion focuses on a number of well-characterized bacterial promoters controlled by one or two species of TFs. We use the results of the thermodynamic model listed in Table 1 of the preceding paper [9] and reproduced as Table 1 in this paper. We make the additional simplifying assumption that the *in vivo* promoters are weak, so that even at full activation, the equilibrium gene expression is still small (e.g., < 10% of the strongest promoters). Indeed for a large number of bacterial promoters, the expression is small in the exponential growth phase when compared to the ribosomal genes, for example, which are fully turned on [12]. In this weak promoter limit, the fold-change in promoter activity (henceforth simply referred to as "fold-change") is given directly by the regulation factor $F_{reg}$ listed in Table 1.

**Simple activation.** The simplest example of activation involves the binding of an induced TF to a single operator site, and the subsequent recruitment of RNAP. This is the case with the *lac* promoter of *E. coli* shown in Fig. 1a (in the absence of the lac repressor) [13,14]. The activating TF is the CRP dimer in complex with cAMP, the inducer molecule. Entry 2 in Table 1 gives the mathematical form of the expected fold-change for this situation with [A]=[CRP$_2$*], $K_A = K$, and Fig. 1b plots its dependence on the induced dimer concentration. The two parameters of the model, the effective *in vivo* dissociation constant $K$ between CRP and the operator, and the enhancement factor *f* that characterizes the degree of stimulation in transcription resulting from operator-bound CRP, are readily revealed in a log-log plot of the relative promoter activity against the cellular concentration of the induced activator, [CRP$_2$*]. As long as the range of [CRP$_2$*] probed is sufficiently broad, one can read the enhancement factor *f* off the graph as the maximal fold-change between full activation at saturating [CRP$_2$*] and basal activity at low [CRP$_2$*]. One can also read off the effective dissociation constant $K$ as the value of [CRP$_2$*] at half-activation. The steepness of the transition region, called the "sensitivity" in the signal transfer literature [15], plays an important role in the function of genetic circuits. Here we quantify transcriptional sensitivity by the log-log slope ($s$) at the mid-point of the transition region. For promoters containing a single operator, $s \leq 1$ and it approaches 1 only for very large $f$'s. In comparison, functions such as amplification, bistability, or spontaneous oscillation all require circuit components to have high sensitivity with $s > 1$ [6,15]

**Cooperative activation**. TFs often have domains that allow for interaction with one another when bound to adjacent operator sites, and this interaction can result in cooperativity in transcriptional activation. The P$_{RM}$ promoter of phage lambda shown in Fig. 2a is such an example [1]. Binding of the dimeric lambda repressor cI to the operator O$_{R2}$ (the "activator" site) stimulates transcription, while binding of cI to the upstream operator O$_{R1}$ (the "helper" site) helps recruit cI to O$_{R2}$. The expected fold-change, Entry 3 in Table 1 with [A]=[H]=[cI$_2$], $K_H = K_{R1}$ and $K_A = K_{R2}$, depends on the affinities $K_{R1}$ and $K_{R2}$ of cI to the two operators, the cooperative interaction $\omega$ between the two operator-bound cI dimers, and the enhancement factor $f$ due to the O$_{R2}$-bound cI. It is shown in the log-log plot of Fig. 2b (thick solid line) as a function of [cI$_2$]/$K_{R2}$.

To quantify the possible role of the auxiliary operator O$_{R1}$, we also plot in Fig. 2b the fold-change for different ratios of $K_{R1}$ and $K_{R2}$. Comparing these curves, it is clear that the auxiliary operator O$_{R1}$ does not change the degree of full activation, given by $f$. The most significant feature of this dual-activator system is perhaps the increase in the log-log slope of the transition region (compared to the extreme cases) for intermediate values of $K_{R2}/K_{R1}$. In fact, for the realistic parameter of $K_{R2}/K_{R1} \approx 25$ (thick solid line), we have a sensitivity of $s \approx 0.93$ which is close to the maximum attainable for this system with its small enhancement factor ($f \approx 11$) and is nearly double the maximum sensitivity $s \approx 0.54$ for the promoter with O$_{R2}$ only (thin solid line). For TFs with larger $f$, this cis-regulatory construct can in principle provide more sensitivity with $s$ approaching 2.

The same cis-regulatory design can be used to implement co-activation, i.e. one of the simplest forms of signal integration, if the two operators are targets of two distinct TF species. An example of this is the variant of *E. coli*'s *melAB* promoter studied by Wade et al [16] (see Fig.

3a), where transcription is stimulated by induced MelR dimer bound to the (weak) proximal operator $O_2$. Meanwhile, CRP bound to the upstream operator $O_1$ helps recruit MelR but does not directly participate in activation. The expected form of the co-dependence is given again by Entry 3 in Table 1, but with [A]=[MelR$_2$*], [H]=[CRP$_2$*] and $K_H = K_1, K_A = K_2$. The fold-change is plotted against the induced CRP concentration on the log-log plot of Fig. 3b for different concentrations of the induced MelR. To better visualize the co-dependence on CRP and MelR, it is useful to plot the fold-change as a 3d plot; see Fig. 3c. The transition region (the yellow band) is clearly dependent on *both* TFs. Consider a simplified situation where CRP and MelR can each take on two possible concentrations, i.e., a pair of "low" and "high" values. Then it is possible to choose the pair of concentrations (e.g., those marked by the 4 open circles) such that the fold-change is large (the green region) only when both concentrations are "high". This mimics a logical AND function of the two inputs [17]. It is also possible to choose the pair of concentrations as marked by the 4 solid circles such that the fold-change is large (the green region) unless both concentrations are "low". The latter mimics a logical OR function. The flexibility of this cis-regulatory scheme makes the shape of the fold-change readily evolvable [18], e.g., between the AND/OR functions, by merely altering the operator sequences which encode the values of $K_1$ and $K_2$.

**Synergistic activation.** An alternative mechanism for co-activation is synergistic or dual activation [19,20,21], where two operator-bound TFs can *simultaneously* contact different subunits of RNAP and activate transcription. This mechanism is limited to TFs that can activate transcription at different locations relative to the core promoter. Prominent examples of such synergistic activation in the bacterial literature [19,20,21,22,23,24,25] all involve the activator CRP since it can recruit RNAP from multiple locations at varying distances upstream of the promoter [8,26].

The synthetic promoter studied by Joung et al. [21] contained two operators, one for cI proximal to the core promoter ($O_2$) and the other for CRP at an upstream operator ($O_1$); see Fig. 4a. The data by Joung et al. supports the model where each operator-bound activator can *independently* interact and recruit RNAP [21]. The expected fold-change is given by Entry 8 in Table 1 (with [A$_1$] = [CRP$_2$*], [A$_2$] = [cI$_2$], $K_{A1} = K_1$, $K_{A2} = K_2$ and $\omega = 1$) and shown in the log-log plot of Fig. 4b as a function of [CRP$_2$*] for various cI concentrations. Note that the dependence of gene expression on [CRP$_2$*] is independent of [cI$_2$], except for an overall vertical shift. This is a reflection of the *multiplicative* nature of independent synergistic activation. An alternative way of visualizing the same result is the 3d plot of Fig. 4c.

In another experiment by Joung et al. [19], the proximal site (O$_2$) was engineered to bind CRP rather than cI (see Fig. 5a, left). An important result of these experiments was that the fold-change with both CRP operators is *larger* than the product of the fold-changes with one operator alone. This is not consistent with the independent recruitment assumption and suggests additional cooperativity ($\omega$). A possible mechanism proposed by Joung et al. is that DNA bending (see Fig. 5a, right) induced by the CRP bound to the proximal operator O2 facilitates the upstream CRP interaction with RNAP, without any direct protein-protein interaction between the two TFs. This cooperative effect can be included in the thermodynamic model as shown in Entry 8 of Table 1 (with [A$_1$] =[A$_2$] = [CRP$_2$*], $K_{A1} = K_1$, $K_{A2} = K_2$ and $\omega > 1$) regardless of the specific molecular mechanism. Like the case of activation by cI, the expression level is most sensitive when the $K$'s for the two binding sites are equal. In Fig. 5b, we plot the expected fold-change with $K_1 = K_2$ and different values of $\omega$. The extra cooperativity increases both the enhancement factor ($\omega \cdot f_1 \cdot f_2$) and the sensitivity ($s$) of the transition region.

**Simple repression.** The simplest example of repression involves the binding of a TF to a single operator site that interferes with the binding of RNAP to the core promoter. This is the case in the truncated *lac* promoter, e.g., lacUV5, which has only the main operator O$_m$ of LacI located closely downstream of the core promoter; see Fig. 6a [27]. The expected fold-change is given by Entry 1 of Table 1, with [R]=[LacI$_4$], $K_R = K_m$ and only one unknown parameter $K_m$ characterizing the effective dissociation constant of the operator O$_m$. Here, it is possible to compute $K_m$ [28] directly from the experimental data of Oehler et al [27] since the cellular concentration of LacI was quantified. In fact, because Oehler et al characterized gene expression at two distinct LacI concentrations, the two data points can be used to check the consistency of the thermodynamic model.

This analysis was performed for the 3 lac operator sequences O1, O2 and O3 studied in [27], with results shown in Fig. 6b. We note that the $K_m$ values obtained, $K_1 \approx 0.22$ nM, $K_2 \approx 2.7$ nM, and $K_3 \approx 110$ nM for the 3 operators, are significantly different from, e.g., the results

$K_1 \approx 10^{-3}$ nM $K_2 \approx 10^{-2}$ nM and $K_3 \approx 0.016$ nM to 1 nM obtained from *in vitro* assays [29,30,31]. These results underscore the fact that the relevant TF-operator binding constant for the thermodynamic model is not given by the *in vitro* measurement (even if the appropriate physiological conditions are used), but must be corrected for by the interaction of the TF with the genomic background [9,32]. Consistent with the theoretical expectation, the ratios of the *K*'s are in reasonable agreement between the *in vivo* and *in vitro* results. We note also that the expected range of promoter activities is much larger than those for the activator-controlled promoters described above. This follows from the strong excluded-volume interaction between the repressor and RNAP, such that more repressor proteins generally lead to stronger repression[1]. In contrast, the sensitivity is limited to $s \leq 1$ with a single operator site.

**Repression by DNA looping**. For the wild-type *lac* promoter, the degree of repression exceeds 1000-fold with merely ~10 repressor molecules in a cell [14]. This is substantially larger than the < 100-fold repression achievable by the best of the truncated promoters (Fig. 6) at the same repressor concentration. The additional repression is facilitated by the stabilization of the $O_m$-bound Lac tetramer which can simultaneously bind to an auxiliary operator $O_a$ through DNA looping (see Fig. 7a). The wildtype *lac* promoter has two such auxiliary operators, O2 located 401 bases downstream and O3 located 92 bases upstream. Here we describe the simpler case studied experimentally by Oehler et al [27], which involves repression and looping only between the main operator $O_m$ and the downstream auxiliary operator O2. The expected fold-change is given by Entry 9 of Table 1, with [R]=[LacI$_4$] and $K_1 = K_m, K_2 = K_a$.

Given that the three *K*'s are already determined (see Fig. 6b), there is only one unknown parameter in this case in the form for the fold-change (Entry 9 of Table 1). It is the effective repressor concentration [*L*] made available via DNA-looping for binding to one of the two operators, *due to* the binding of a repressor to the other operator. Oehler et al [27] did experiments with the main operator $O_m$ as one of the 3 operator sequences (O1, O2, O3), each for two concentrations of LacI. The results of all 6 experiments are consistently described by the expected fold-changes according to the thermodynamic model (see Fig. 7b), with $[L] \approx 660$ nM [28].

---

[1] Not discussed here is a lower plateau of promoter activity set by promoter leakage.

Quantitatively, the strong repression effect (compare Fig. 6b and Fig. 7b) results directly from the large value of [$L$] generated by DNA looping, which effectively amplifies one operator-bound repressor 660 fold. This enhancement of the local repressor concentration is due to the *linkage* between $O_m$ and $O_a$, as already described qualitatively in Refs. [27,33]. Intuitively, once a LacI tetramer binds to one of the two operators, say $O_m$, it is available within a small volume for binding to $O_a$. The actual value of [$L$] is clearly dependent on the spacing between the two operators, as well as the energetics of bending the DNA backbone. We have deduced the dependence of [$L$] on operator spacing (shown in Fig. 7d) by analyzing the data of Müller et al [34], who measured the fold-changes in repression for promoter constructs with different spacing between the main and auxiliary operators; see Fig. 7c. In Fig. 7c, we also show the *predicted* transcriptional fold-changes for the same constructs of Muller et al [34] but at different LacI concentrations.

**Cooperative repression**. Interaction between the TFs can also promote the sensitivity in transcriptional repression. The $P_R$ promoter, which controls the expression of *cro* in phage lambda (illustrated in Fig. 8a), is a good example of this mode of repression [1]. When bound to either the operator $O_{R1}$ or $O_{R2}$, the lambda repressor cI blocks the access of RNAP to the core promoter, thereby repressing transcription. The combined effect of two repressive operators, reinforced by the cooperative interaction between the operator-bound cI's, results in further repression. The expected form of fold-change is given by Entry 6 in Table 1 ([$R_1$]=[$R_2$]= [$cI_2$] and $K_{R1} = K_{R2}$, $K_{R2} = K_{R1}$) and plotted in Fig. 8b. As with the case of cooperative activation (Fig. 2), maximum log-log slope (i.e. sensitivity) in repression is the largest when $K_1$ and $K_2$ are comparable. Similar schemes have been generalized for co-repression by two species of repressors [35,36,37], and can be used to mimic the logical NAND function [17].

Cooperativity in repression in fact does *not* require direct interaction between the repressor molecules. An example is the $P_{LtetO-1}$ promoter [38], which contains two operators of TetR; see Fig. 8c. The expected form of the fold-change is given by Entry 5 in Table 1 with [$R_1$] =[$R_2$] = [TetR$_2$*], and $K_{R1} = K_1$, $K_{R2} = K_2$. Because the occupation of either operator is sufficient to block RNAP from the core promoter, it follows that the fold-changes (not shown) are almost identical to those of Fig. 8b even though the TetR dimers do not interact [39]. We expect that a similar construct where the two operators are targets of different, non-interacting TFs can implement co-repression. Comparing the activating and repressive modes of transcription control, we find repressive control to be advantageous because (i) high sensitivity can be generated by TFs without the need of TF-TF interaction, and (ii) fold changes are not limited by the magnitude of the (typically weak) TF-RNAP interaction [40].

**Phenomenological Model of Transcription Control**.

The mathematical description for the different activation and repression mechanisms discussed above can be summarized by very simple forms. For a single TF species with up to two operators in the cis-regulatory region, all of the fold-changes described in Table 1 can be compactly represented by the general form

$$F_{reg}([TF]) = \frac{1 + a_1[TF] + a_2[TF]^2}{1 + b_1[TF] + b_2[TF]^2}. \tag{1}$$

Similarly for co-regulation by two TFs with cellular concentrations $[TF_1]$ and $[TF_2]$, and for no more than one operator each in the regulatory region, the fold-change has the form

$$F_{reg}([TF_1],[TF_2]) = \frac{1 + a_{1,0}[TF_1] + a_{0,1}[TF_2] + a_{1,1}[TF_1] \cdot [TF_2]}{1 + b_{1,0}[TF_1] + b_{0,1}[TF_2] + b_{1,1}[TF_1] \cdot [TF_2]}. \tag{2}$$

The general forms (1) and (2) include many possible mechanisms of activation and repression not discussed above. If 3 binding sites for the TF are involved in the regulatory process, then Eq. (1) or (2) would be generalized to the ratio of $3^{rd}$ degree polynomials of the $[TF]$'s.

The above analysis indicates that by quantitatively measuring the fold-change as a function of the activated TF concentration(s), we can achieve two important goals: (i) By fitting experimental results to an expression such as (1) or (2), one would obtain a quantitative characterization of the promoter at all TF concentrations by only a few (e.g., 4 or 6) parameters. This can be done regardless of the validity of the thermodynamic model itself. As discussed previously, the compact description will facilitate quantitative higher-level study of gene circuits. (ii) By comparing the values of these parameters to the expected forms according to the thermodynamic model (e.g., Table 1), one can generate hypotheses on the likely mechanisms of transcriptional control for further experiments. Thus the form of the fold-change in gene expression itself can be an effective diagnostic tool to distinguish subtle mechanisms of transcriptional control.

## Conclusion

We have illustrated a variety of promoter activities implemented in different cis-regulatory designs. Also illustrated are important functional differences (e.g., in transcriptional cooperativity, and in the nature of combinatorial control) among promoters characterized by different parameters of the same cis-regulatory construct. These differences often cannot be discriminated by the qualitative characterization of promoter activity predominantly practiced in molecular biology today (e.g., fold-change in gene expression due to deletion of a regulatory protein). Instead, they call for more quantitative characterization, particularly the quantification of the TF concentrations (or their relative concentrations) controlling promoter activity. The reward of quantitative characterization includes a compact phenomenological description of promoter activity for higher-level analysis and the elucidation of unknown mechanisms of transcriptional control.

## Acknowledgements

We are grateful to Steve Busby, Ann Hochschild, Bill Loomis, Mark Ptashne, Milton Saier Jr, and Jon Widom for discussions and comments. This research is supported by the NIH Director's Pioneer Award (RP), NSF through grants 9984471, 0403997 (JK), and 0211308, 0216576, 0225630 (TH, TK). JK is a Cottrell Scholar of Research Corporation. UG acknowledges an 'Emmy Noether' research grant from the DFG.

| Case | Regulation Factor ($F_{reg}$) | |
|---|---|---|
| **1. Simple repressor** 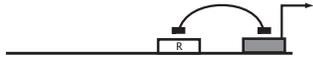 | $(1+r)^{-1}$ | $\left(1+\frac{[R]}{K_R}\right)^{-1}$ |
| **2. Simple activator** 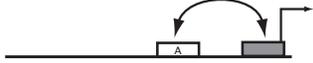 | $\dfrac{1+ae^{-\frac{\varepsilon_{ap}}{k_BT}}}{1+a}$ | $\dfrac{1+\frac{[A]}{K_A}f}{1+\frac{[A]}{K_A}}$ |
| **3. Activator recruited by a helper (H)** 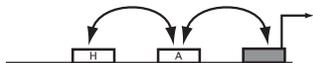 | $\dfrac{1+a\frac{1+he^{-\frac{\varepsilon_{hq}}{k_BT}}}{1+h}e^{-\frac{\varepsilon_{ap}}{k_BT}}}{1+a\frac{1+he^{-\frac{\varepsilon_{hq}}{k_BT}}}{1+h}}$ | $\dfrac{1+\frac{[H]}{K_H}+\frac{[A]}{K_A}f+\frac{[A]}{K_A}\frac{[H]}{K_H}f\omega}{1+\frac{[H]}{K_H}+\frac{[A]}{K_A}+\frac{[A]}{K_A}\frac{[H]}{K_H}\omega}$ |
| **4. Repressor recruited by a helper (H)** 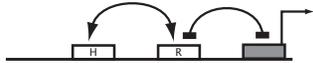 | $\left(1+\frac{1+he^{-\frac{\varepsilon_{hq}}{k_BT}}}{1+h}r\right)^{-1}$ | $\dfrac{1+\frac{[H]}{K_H}}{1+\frac{[H]}{K_H}+\frac{[R]}{K_R}+\frac{[R]}{K_R}\frac{[H]}{K_H}\omega}$ |
| **5. Dual repressors** 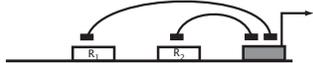 | $(1+r_1)^{-1}(1+r_2)^{-1}$ | $\left(1+\frac{[R_1]}{K_{R_1}}\right)^{-1}\left(1+\frac{[R_2]}{K_{R_2}}\right)^{-1}$ |
| **6. Dual repressors interacting** 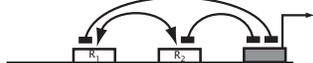 | $\left(1+r_1+r_2+r_1r_2e^{-\frac{\varepsilon_{r_1r_2}}{k_BT}}\right)^{-1}$ | $\left(1+\frac{[R_1]}{K_{R_1}}+\frac{[R_2]}{K_{R_2}}+\frac{[R_1]}{K_{R_1}}\frac{[R_2]}{K_{R_2}}\omega\right)^{-1}$ |
| **7. Dual activators interacting** 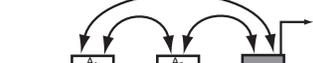 | $\dfrac{1+a_1e^{-\frac{\varepsilon_{a_1p}}{k_BT}}+a_2e^{-\frac{\varepsilon_{a_2p}}{k_BT}}+a_1a_2e^{-\frac{\varepsilon_{a_1p}+\varepsilon_{a_2p}+\varepsilon_{a_1a_2}}{k_BT}}}{1+a_1+a_2+a_1a_2e^{-\frac{\varepsilon_{a_1a_2}}{k_BT}}}$ | $\dfrac{1+\frac{[A_1]}{K_{A_1}}f_1+\frac{[A_2]}{K_{A_2}}f_2+\frac{[A_1]}{K_{A_1}}\frac{[A_2]}{K_{A_2}}f_1f_2\omega}{1+\frac{[A_1]}{K_{A_1}}+\frac{[A_2]}{K_{A_2}}+\frac{[A_1]}{K_{A_1}}\frac{[A_2]}{K_{A_2}}\omega}$ |
| **8. Dual activators cooperating via looping** 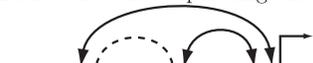 | $\dfrac{1+a_1e^{-\frac{\varepsilon_{a_1p}}{k_BT}}+a_2e^{-\frac{\varepsilon_{a_2p}}{k_BT}}+a_1a_2e^{-\frac{\varepsilon_{a_1p}+\varepsilon_{a_2p}+F_{loop}}{k_BT}}}{(1+a_1)(1+a_2)}$ | $\dfrac{1+\frac{[A_1]}{K_{A_1}}f_1+\frac{[A_2]}{K_{A_2}}f_2+\frac{[A_1]}{K_{A_1}}\frac{[A_2]}{K_{A_2}}f_1f_2\omega}{(1+\frac{[A_2]}{K_{A_2}})(1+\frac{[A_1]}{K_{A_1}})}$ |
| **9. Repressor with two DNA binding units and DNA looping** 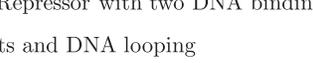 | $\left(1+r_1+\frac{r_1}{1+r_2}e^{-\frac{\Delta\varepsilon_{rd_2}+F_{loop}}{k_BT}}\right)^{-1}$ | $\dfrac{1+\frac{[R]}{K_2}}{\left(1+\frac{[R]}{K_1}\right)\left(1+\frac{[R]}{K_2}\right)+\frac{[R]\cdot[L]}{K_1\cdot K_2}}$ |
| **10. N nonoverlapping activators and/or repressors acting independently on RNAp** 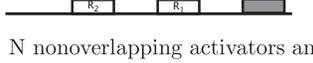 | $F_{reg1}\cdot F_{reg2}\cdot\ldots\cdot F_{regN}$ | $F_{reg1}\cdot F_{reg2}\cdot\ldots\cdot F_{regN}$ |

**Table 1** (reproduced from [9]): Regulation factors for a number of different regulatory motifs. The second column gives the regulation factor in terms of the number of transcription factors (TFs) in the cell and their binding energies, while the third column provides a translation of the regulation factor into the language of concentrations and equilibrium dissociation constants. For an arbitrary TF we introduce the following symbols: in the second column $x$ is the combination $\dfrac{X}{N_{NS}}e^{-\Delta\varepsilon_{xd}/k_BT}$, while $[X]$ in the third column denotes the concentration of transcription factor X. $K_X=[X]/x$ is the effective equilibrium dissociation constant of the TF and its operator sequence on the DNA. Furthermore, in the third column we introduce $f=e^{-\varepsilon_{xp}/k_BT}$ for the "glue-like" interaction of a TF and RNAP, and for $\omega=e^{-\varepsilon_{x_1x_2}/k_BT}$ the interaction between two TFs. In entries 8 and 10, $F_{loop}$ is the free energy of looping out DNA, $\omega$ in 8 is defined as $e^{-F_{loop}/k_BT}$, while [L] in 9 is the combination $\dfrac{N_{NS}}{V_{cell}} e^{-F_{loop}/k_BT}$; $V_{cell}$ is the volume of the cell.

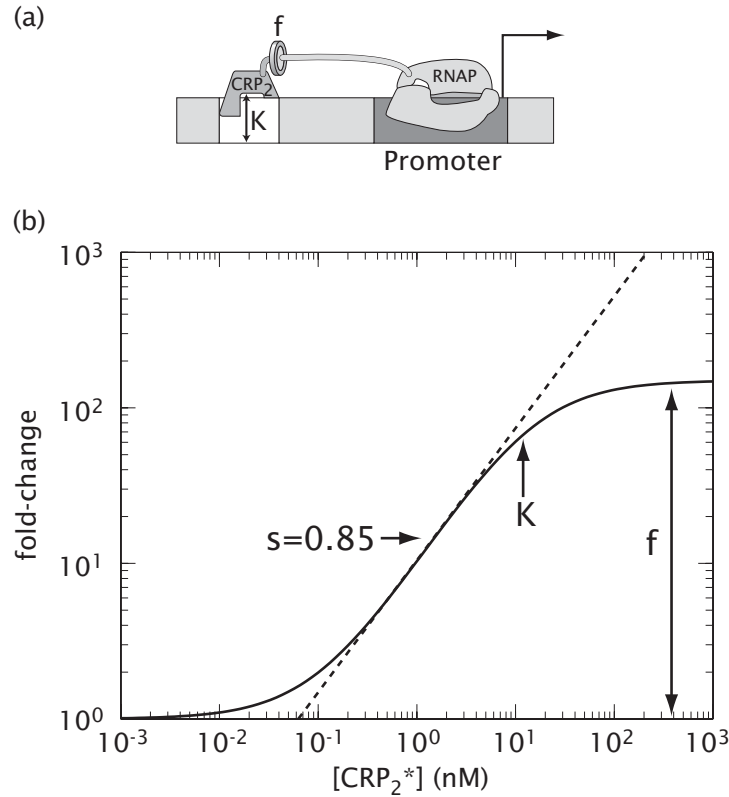

**Figure 1:** (a) Cis-regulatory architecture for transcriptional activation involving a single CRP operator, as found in the *lac* operon. The white box denotes the operator site and the dark box corresponds to the promoter. The DNA-binding affinity of the transcription factor for its operator is described by the *in vivo* dissociation constant *K*, which is the TF concentration at which the operator occupancy is half-maximal. The activator recruits RNAP through protein-protein interactions (schematically drawn as interacting protein subunits). (b) Log-log plot of the fold-change in gene expression as a function of the induced CRP concentration, $[CRP_2^*]$. The maximum log-log slope in the transition region, which is defined as the sensitivity (s), is highlighted with the dashed line and is equal to 0.85. This plot was generated using $K = 15$ nM, $f = 150$. These parameter values were estimated from experiments similar to those of Setty et al. [10], who measured beta-galactosidase activity as a function of extra-cellular cAMP concentration in *E. coli* MG1655 cells, but with the additional deletion of the *cyaA* gene which encodes adenyl cyclase (T. Kuhlman and T. Hwa, unpublished data). The estimated value of the effective dissociation constant *K* is dependent on the literature values for a number of biochemical parameters concerning cAMP binding and transport, and is not expected to be accurate to within a factor of 2. (For comparison, previous *in vitro* measurement of the CRP-operator affinity has ranged from 0.001nM to 50nM depending on the ionic strength of the assay [41,42,43]).

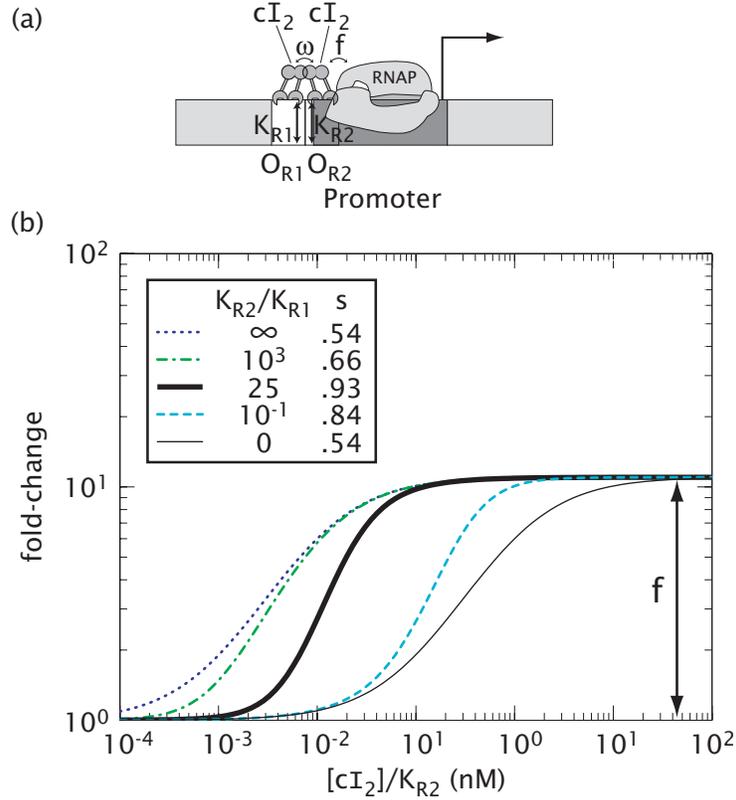

**Figure 2:** (a) Cis-regulatory architecture for cooperative transcriptional activation in phage lambda $P_{RM}$ promoter. Here we are considering $P_{RM}$ alone without the upstream $P_L$ region which affects $P_{RM}$ activity through DNA looping [44]. We also neglect the operator $O_{R3}$, which has very weak affinity to cI in the absence of $P_L$ [44]. The white boxes denote the operator sites $O_{R1}$, $O_{R2}$ and the dark box corresponds to the promoter. The DNA-binding affinity of $cI_2$ for $O_{R1}$ and $O_{R2}$ is described by the dissociation constants $K_{R1}$ and $K_{R2}$, respectively. The activator recruits RNAP and cI dimers interact with one another through intimate, cooperative interactions, both of which are indicated by overlapping protein-protein domains. (b) Log-log plot of the fold-change in gene expression as a function of $cI_2$ concentration for different ratios of $K_{R2}/K_{R1}$. The maximum log-log slopes ($s$) for the different curves are listed in the legend. The promoter with $K_{R2}/K_{R1} = 0$ corresponds to a deletion of $O_{R1}$, and the regulation function for this case (thin solid line) is identical to the single operator case shown in Fig. 1. If this promoter has a very small $K_{R1}$ (i.e. strong $O_{R1}$), then the onset of full activation will be shifted to smaller cI concentrations (dotted line). The latter corresponds effectively to a stronger $O_{R2}$ site, with dissociation constant $K_{R2}/\omega$. The transition region is the steepest when the two $K$'s are comparable. These plot are generated using $f \approx 11$ [45] and $\omega \approx 100$ [46] as extracted from *in vitro* biochemical studies. The absolute *in vivo* values of the $K$'s are not known (which is why the concentration is expressed in terms of $[cI_2]/K_{R2}$). However, the ratio $K_{R2}/K_{R1} \approx 25$ (thick solid line) can be deduced from the *in vitro* results [46].

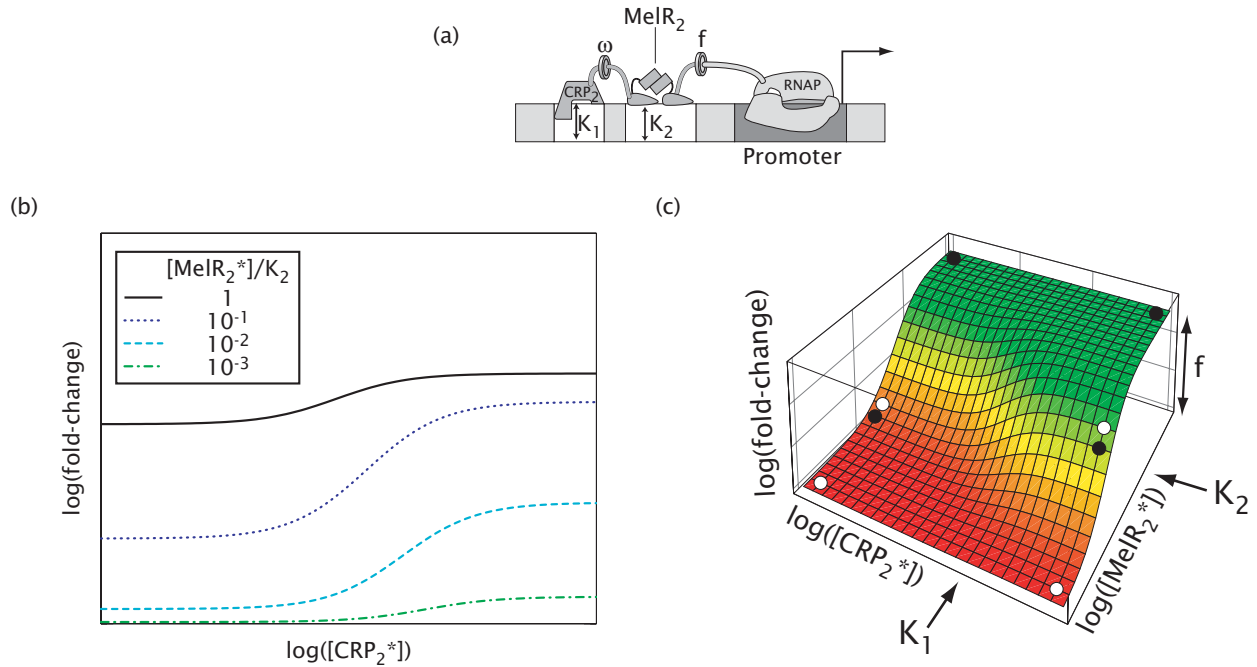

**Figure 3:** (a) Cis-regulatory construct for co-activation by CRP and MelR. The figure shows the truncated JK15 version of *melAB* promoter studied by Wade et al [16]. The full *melAB* promoter is more complicated due to the presence of multiple MelR operators. However, the co-activation pattern is similar to that of JK15 discussed here. The white boxes denote the operator sites $O_1$, $O_2$ and the dark box corresponds to the promoter. The DNA-binding affinity of $CRP_2$ for $O_1$ and $MelR_2$ for $O_2$ is described by the dissociation constant $K_1$ and $K_2$, respectively. MelR can recruit RNAP (drawn with protein-protein contacts) and cooperative interaction between $MelR_2$ and $CRP_2$ is indicated by interacting protein subunits. (b) Log-log plot of the fold-change in gene expression as a function of activated MelR dimer concentration $[MelR_2^*]$ for different activated CRP dimer concentrations $[CRP_2^*]$. Since none of the parameters $f$, $\omega$, and $K$'s has been determined experimentally, the scales of the plot can only be expressed relative to these parameters. Nevertheless, the plot reveals important qualitative predictions by the thermodynamic model, e.g. the dependence of the maximal Crp-dependent fold-change on the MelR concentration. (c) 3d log-log plot of the fold-change in gene expression as a function of both of $CRP_2$ and $MelR_2$. For different choices of "high" and "low" concentration (the four combinations of "high/low" for these two TFs form a rectangle), the same *melAB* promoter can serve as an OR-function (solid circles) or an AND-function (open circles).

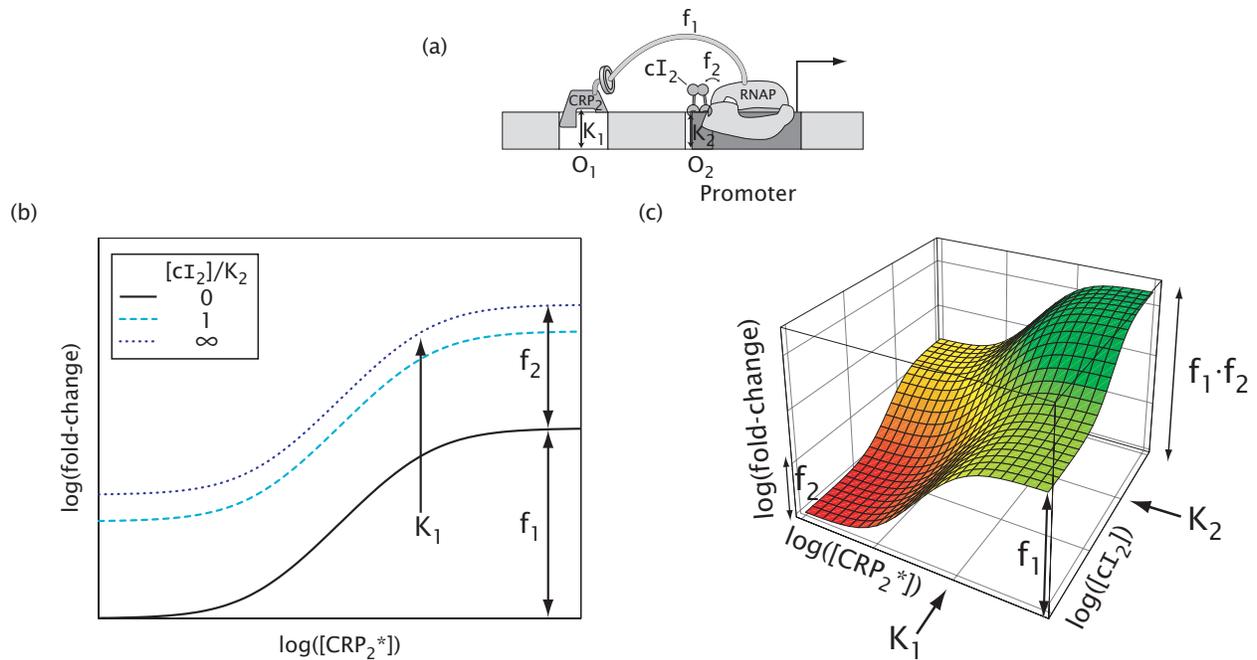

**Figure 4:** (a) Cis-regulatory architecture for synergistic co-activation in synthetic promoters [21]. The white boxes denote the operator sites $O_1, O_2$ and the dark box corresponds to the promoter. The DNA-binding affinity of $CRP_2$ for $O_1$ and $cI_2$ for $O_2$ is described by the dissociation constant $K_1$ and $K_2$, respectively. Either activator can recruit RNAP independently at different strengths $f_1$, $f_2$ (as shown with interacting protein-protein subunits). (b) Log-log plot of the fold-change in gene expression as a function of $[CRP_2^*]$ for different concentrations of $[cI_2]$. (c) 3d log-log plot of the fold-change in gene expression as a function of both $CRP_2$ and $cI_2$. Note that on log scale, the product appears as an additive shift.

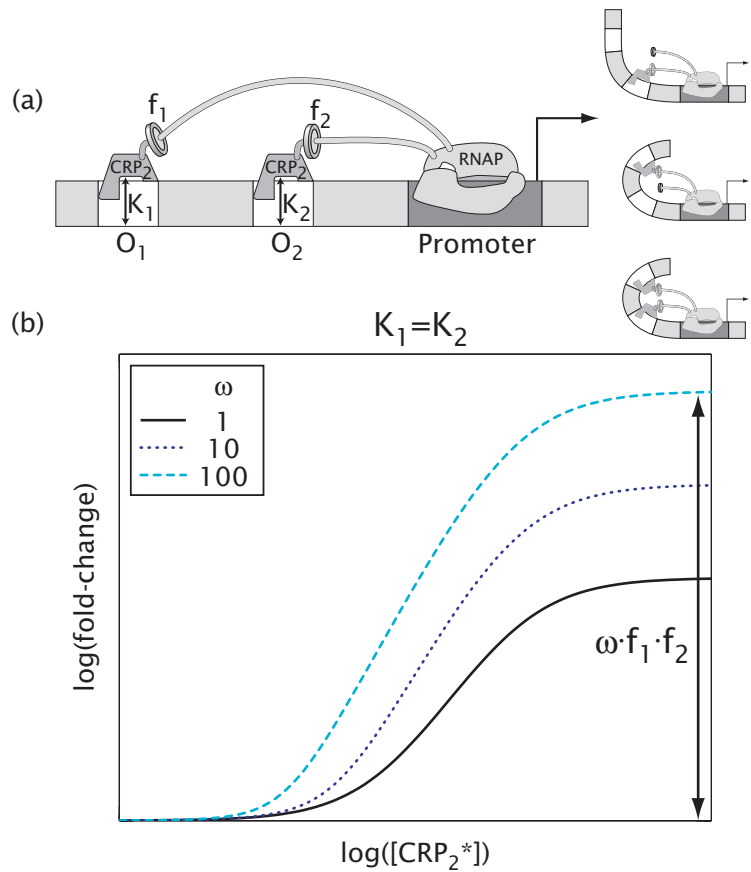

**Figure 5:** (a) To the left is the cis-regulatory architecture for synergistic activation by the same TF in synthetic promoters [19]. The white boxes denote the operator sites $O_1$, $O_2$ and the dark box corresponds to the promoter. The DNA-binding affinity of $CRP_2$ for $O_1$ and $O_2$ is described by the dissociation constants $K_1$ and $K_2$, respectively. Activators at each operator can recruit RNAP independently at different strengths $f_1$, $f_2$ (as shown with interacting protein-protein subunits). As illustrated to the right, the binding of CRP to proximal $O_2$ bends DNA and facilitates the "bent" interaction of RNAP to CRP bound at upstream $O_1$ (b) Log-log plot of the fold-change in gene expression as a function of $[CRP^*_2]$ for equal dissociation constants ($K_1=K_2$). We have included the additional cooperativity $\omega$ that can occur when the binding of CRP to $O_1$ promotes the interaction of RNAP to CRP bound at $O_2$. The maximal fold-change is $\omega \cdot f_1 \cdot f_2$.

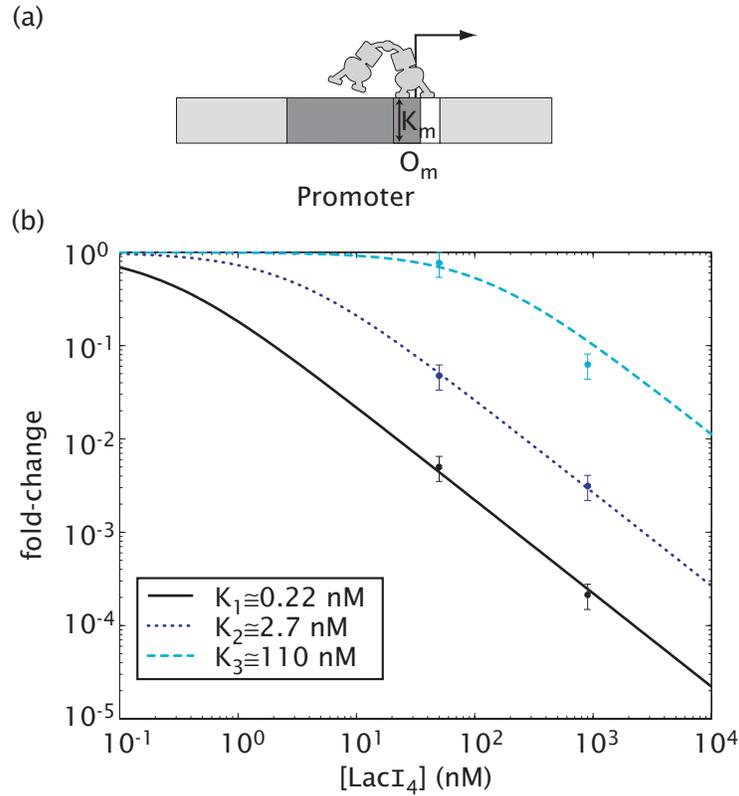

**Figure 6:** (a) Cis-regulatory structure of the truncated *lac* promoter, with the main operator $O_m$ (white box) located closely downstream of the core promoter (dark box). Repressor bound at $O_m$ will block RNAP binding to the promoter, as denoted by the overlap. The DNA-binding affinity of $LacI_4$ for $O_m$ is described by the dissociation constant $K_m$. (b) Log-log plot of the fold-change in gene expression as a function of $LacI_4$. Here, the repressor concentration shown on the horizontal axis refers to the cellular LacI tetramers in the absence of inducers. The experiments of Oehler et al.[27] used the operator sequences O1, O2, O3 at position $O_m$ and measured fold-repression at two different LacI concentrations (50nM and 900nM); the data are shown as circles. The expected form of the fold-changes are plotted as the solid, dotted and dashed lines as indicated in the legend. The value of $K_m$ for each curve (see legend) is determined by fitting one of the two data points. The fact that the other data point lies closely on the curve supports the applicability of the thermodynamic model to this promoter.

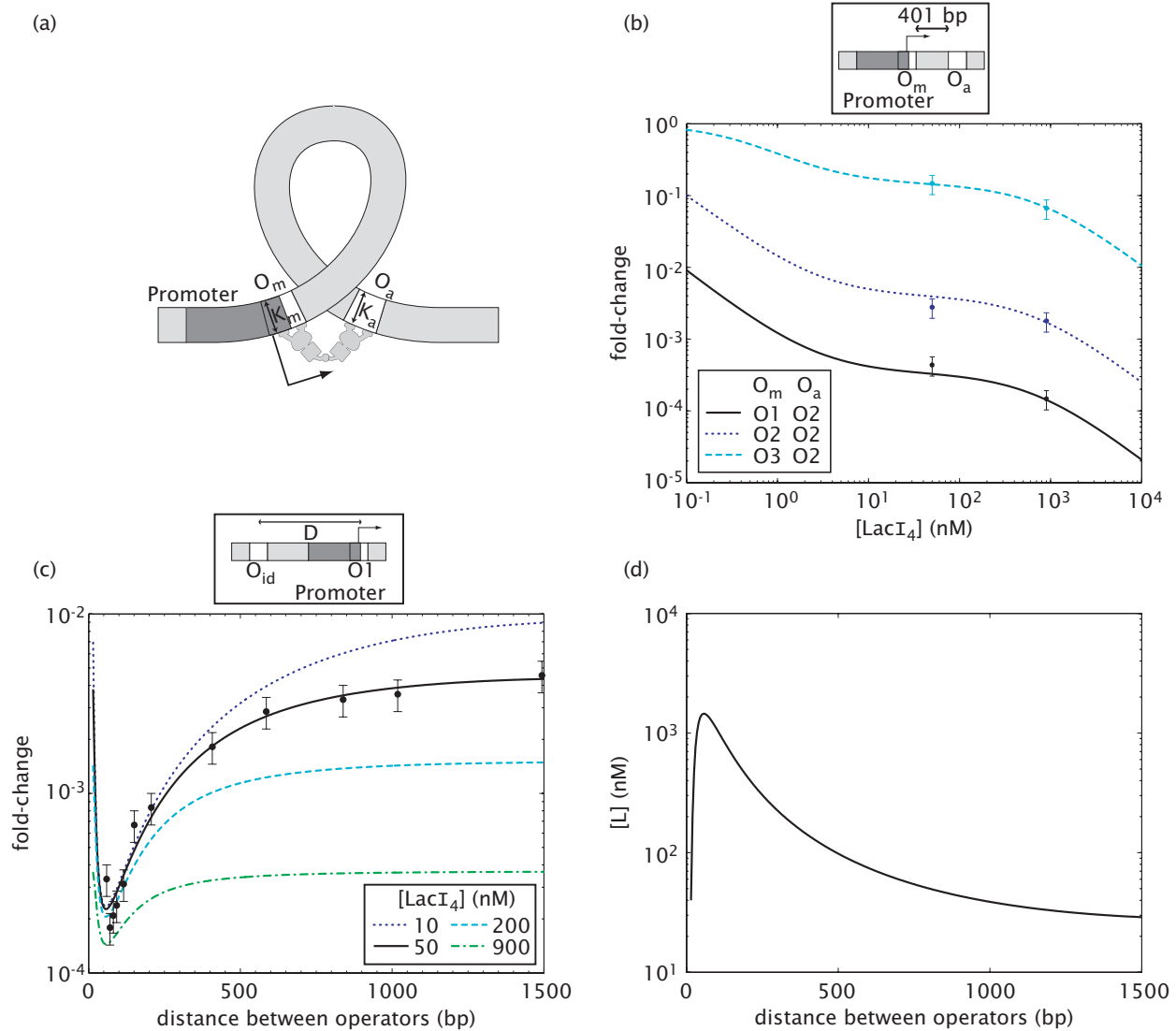

**Figure 7:** (a) Cis-regulatory layout for looping and repression in the *lac* promoter experiments of Oehler et al [27]. White boxes are operators and the dark box is the promoter. LacI tetramer bound at the main operator $O_m$ interferes with RNAP binding to the promoter. This binding is further stabilized if the other two legs of the tetramer bind at $O_a$ through DNA-looping. (b) Log-log plot of the fold-change in gene expression as a function of $LacI_4$ concentration for different constructs where $O_m$ is replaced by O1, O2, or O3 and $O_a$ is O2. The curves are generated by plotting entry 9 of Table 1 using the appropriate dissociation constants shown in Fig. 6, for each pair of operators involved. Note that the 6 data points (shown with circles) can all be brought into agreement with the expected form (the lines) by the choice of a single parameter, the available $LacI_4$ concentration $[L]$ due to looping. The best-fit value obtained is $[L] \approx 660$ nM. (c) Log-linear plot of the transcriptional fold-change as a function of distance $D$ between O1 (located at position $O_m$) and an auxiliary operator $O_{id}$ located upstream of the promoter, for various repressor concentrations. The data of [34] (filled circles) are fitted to the transcriptional fold-changes expected for looping (solid line) using $[LacI_4] = 50$ nM and values of $K_1 \approx 0.27$ nM

and $K_{id} \approx 0.05$ nM determined from the data of [27]. The fitting function is the dependence of the available concentration due to looping, $[L]$, on the operator spacing $D$. We use the form $[L] = \exp(-a/D - b \cdot \ln D + c \cdot D + e)$ motivated by the worm-like chain model of DNA bending [47]. The other lines correspond to the predicted gene expression of the same constructs at different LacI concentrations as indicated in the legend. (d) Log-linear plot of $[L]$ vs $D$ obtained from the fit described in (d), with $a = 140.6$, $b = 2.52$, $c = 1.4 \times 10^{-3}$, $e = 19.9$.

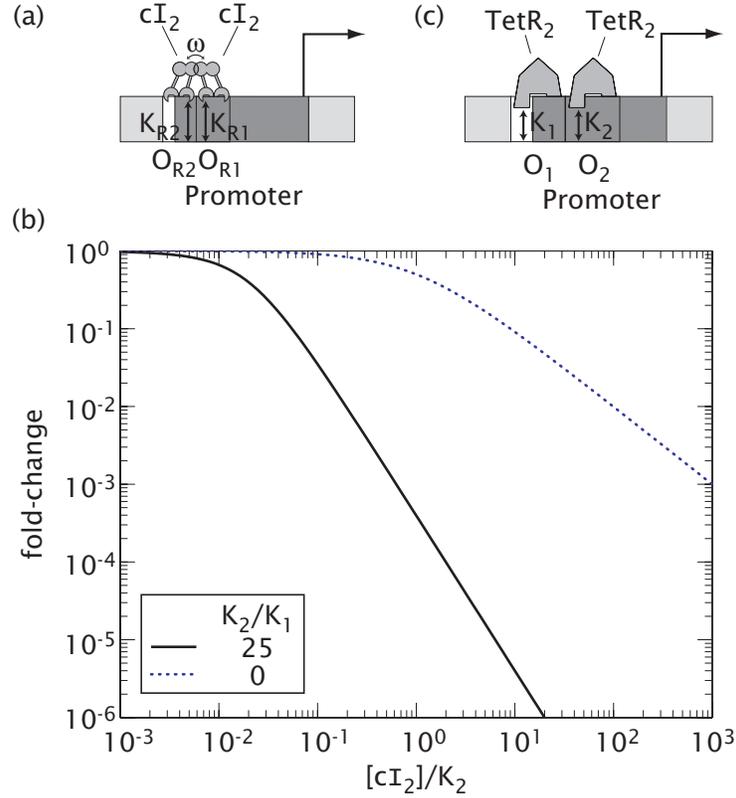

**Figure 8:** (a) Cis-regulatory architecture for cooperative transcriptional repression in phage lambda $P_R$ promoter. The white boxes denote the operator sites $O_{R1}$, $O_{R2}$ and the dark box corresponds to the promoter. Repression is indicated by the overlap between the promoter and operator. The cooperative interaction between bound $cI_2$ at operators $O_{R1}$ and $O_{R2}$ is given by $\omega$ (protein-protein contacts) (b) Log-log plot of the fold-change in gene expression as a function of $cI_2$ concentration for two different values of $K_{R2}/K_{R1}$. At high repressor concentrations, the maximum log-log slope (s) for all the curves is equal to 2 with the exception of $K_{R2}/K_{R1}=0$ (i.e. deletion of $O_{R1}$) where the maximum log-log slope is equal to 1. The latter case corresponds to a single repressive site, $O_{R2}$ (see Fig. 6). This plot was generated using $\omega \approx 100$, and $K_{R2}/K_{R1} \approx 25$ extracted from *in vitro* biochemical studies [46]. The absolute *in vivo* value of the $K$'s are unknown, which is why our concentration is expressed in terms of $[cI_2]/K_{R2}$. (c) Cis-regulatory architecture for transcription repression in $P_{LtetO-1}$ promoter engineered by Lutz & Bujard [38]. Note that there is no cooperative interaction between the TetR dimers. The log-log plot of fold-change of $P_{LtetO-1}$ promoter is similar to that of phage lambda $P_R$ with a maximum log-log slope equal to 2.